\documentclass[aps,twocolumn,showpacs,preprintnumbers]{revtex4}
\usepackage{amssymb}
\usepackage{amsmath}
\usepackage{t1enc}
\usepackage{graphicx}
\usepackage{psfrag}

\begin{document}

\newcommand{\densop}[1]{\hat{\rho}_{#1}}
\newcommand{\slowdensop}[1]{\tilde{\rho}_{#1}}
\newcommand{\dens}[1]{\rho_{#1}}
\newcommand{\mean}[1]{\left<#1\right>}
\newcommand{\ket}[1]{\left|#1\right>}
\newcommand{\bra}[1]{\left<#1\right|}
\renewcommand{\H}[0]{\hat{H}}
\newcommand{\F}[1]{\hat{F}_{#1}}
\renewcommand{\vec}[1]{\mathbf{#1}}
\renewcommand{\Re}[0]{\mathrm{Re}}
\renewcommand{\Im}[0]{\mathrm{Im}}
\newcommand{\exppl}[0]{e^{i\omega t}}
\newcommand{\expml}[0]{e^{-i\omega t}}
\newcommand{\commutator}[2]{\left[#1,#2\right]}

\title{Characterizing the spin state of an atomic ensemble \\  using the
       magneto-optical resonance method.}
\author{B. Julsgaard, J. Sherson, J. L. S\o rensen}
\affiliation{QUANTOP - Danish Quantum Optics Center\\
Institute of Physics and Astronomy, University of Aarhus, 8000 Aarhus,
Denmark}

\author{E. S. Polzik}
\email{polzik@nbi.dk}
\affiliation{QUANTOP - Danish Quantum Optics Center\\
Niels Bohr Institute, 2100 Copenhagen \O,
Denmark}

\date{\today }

\begin{abstract}
  Quantum information protocols utilizing atomic ensembles require
  preparation of a coherent spin state (CSS) of the ensemble as an
  important starting point.  We investigate the magneto-optical
  resonance method for characterizing a spin state of cesium atoms in
  a paraffin coated vapor cell. Atoms in a constant magnetic field are
  subject to an off-resonant laser beam and an RF magnetic field. The
  spectrum of the Zeeman sub-levels, in particular the weak quadratic
  Zeeman effect, enables us to measure the spin orientation, the
  number of atoms, and the transverse spin coherence time. Notably the
  use of 894nm pumping light on the $D1$-line, ensuring the state
  $F=4$, $m_F=4$ to be a dark state, helps us to achieve spin
  orientation of better than 98\%. Hence we can establish a CSS with
  high accuracy which is critical for the analysis of the entangled
  states of atoms.
\end{abstract}

\maketitle

%\preprint{}
%

%\homepage{http://www.ifa.au.dk/amo/qoptics/qoptics.htm}

\section{Introduction}
The spin state of ensembles of alkaline atoms have been studied for a
long time in many different contexts, e.g.~sensitive magnetic field
measurements \cite{alexandrov:96}, frequency standards
\cite{salomon:01}, and recently within the field of quantum information. 
Coherent transfer of states of the
electromagnetic field to atomic spin states with the aid of
electromagnetically induced transparency has been demonstrated
\cite{phillips:01,liu:01}. A sample of cesium atoms has been spin
squeezed \cite{kuzmich:00a}, two samples of cesium atoms have been
entangled \cite{julsgaard:01}, and in \cite{schori:02,julsgaard:03}
the sensitivity of atomic spin states to quantum fluctuations of the
electromagnetic field was demonstrated.  Some of these experiments
require the preparation of a coherent spin state, CSS, (e.g.~all atoms
pumped into the magnetic substate $F=4$, $m_F=4$ in the cesium ground
state). 

This paper provides a detailed report on the creation and
characterization of the coherent spin state used in the recent
experiments on entanglement and quantum memory
\cite{julsgaard:01,schori:02,julsgaard:03}. The importance of the CSS
of a macroscopic ensemble is primarily in that the transverse spin
components of this state are in the minimum uncertainty state and thus
provide a necessary starting point for observing quantum effects. For
example, in order to apply the necessary and sufficient conditions of
the entanglement and spin squeezing \cite{duan:00b,wineland:92} one has to
know the variance of the CSS. Unlike in case of the shot noise of
light, atomic coherent state noise cannot be easily established via
independent measurements. It therefore has to be determined
experimentally in each case.

In this paper we report measurements of the orientation, the coherence
time, and the number of atoms in a spin ensemble in one selected
hyperfine ground state (e.g.~the $F=4$ states in cesium).  This is
done by inducing weak transitions among the Zeeman sublevels by
radio-frequency magnetic fields. Other similar probing methods have
been demonstrated in \cite{avila:87} where microwave magnetic fields
on the hyperfine transition are used to characterize the efficiency of
optical pumping into e.g.~the $F=3, m_F=0$ state. Also, fluorescence
from atoms excited by the pumping process can give information about
the excited spin state which again hints on the ground spin state
\cite{fischer:82}. In the literature, the spin state of atoms is often
modelled by numerical solutions to rate equations. We make a different
approach by tailoring simple ad hoc models to describe the main
features of the spin state.

The paper is arranged as follows. In section~\ref{sec:theor-descr} we
describe the ground spin state of our atoms theoretically. This
includes an introduction to the notation and the particular physical
system used in the experiment~\ref{sec:descr-spin-state}, deriving
equations of motion~\ref{sec:solut-equat-moti}, and modelling the
distribution and coherence time of atoms among Zeeman
sublevels~\ref{sec:modelling-spin-state}. In the experimental
section~\ref{sec:experiment} we describe the actual
setup~\ref{sec:experimental-setup}, discuss the conditions for
resolving the quadratic Zeeman effect~\ref{sec:resolv-diff-zeem}, and
demonstrate that our models actually describe the
experiments~\ref{sec:conf-spin-model}. Much of the emphasis is put on
the quadratic Zeeman effect and the ability to resolve this
spectroscopically. However, in some experimental conditions this is
not the case and we discuss how we can employ our techniques in a
non-resolved regime~\ref{sec:unresolved} and also in a regime with
pulsed lasers~\ref{pulsed_exp}.  In
appendix~\ref{sec:quadr-zeem-effect} we review the quadratic Zeeman
effect.

\section{Theoretical description}
\label{sec:theor-descr}
In this section, we introduce the physics and notation which will
enable us to understand how magneto-optical resonance can be used to
characterize the spin state of an atomic gas sample.

\begin{figure}
  \psfrag{name}{$\begin{matrix} 
                 \Gamma_{\mathrm{com}} = 9.4\mathrm{Hz} \\
                 \Gamma_{\mathrm{pump}} = 0.0\mathrm{Hz} \\
                 J_z = 0.122\mathrm{[a.u.]} \\
                 p = 0.346
               \end{matrix}$}
\includegraphics{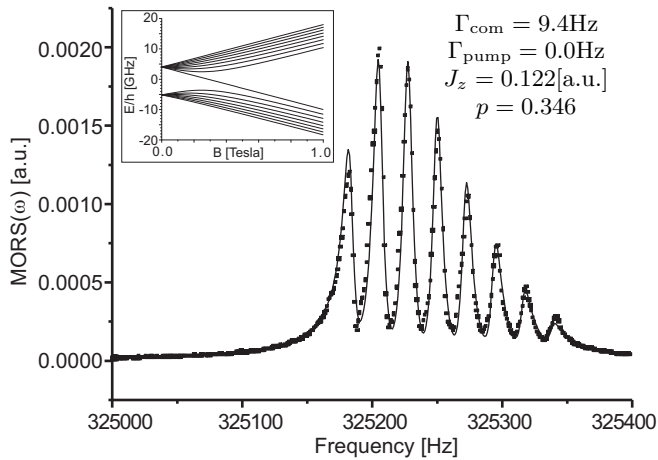}
\caption{An experimental spectrum (dots) of magnetic transitions among 
  the nine sublevels of the $F=4$ ground state in cesium. An external
  RF-magnetic field (the frequency of which is shown on the abscissa)
  modulates the spin state. The recorded response, called the
  magneto-optical resonance signal (MORS), is shown on the ordinate
  axis. The solid line is a fit to a model to be developed in
  section~\ref{sec:modelling-spin-state}, where the parameters and the
  definition of MORS will also be carefully explained. The many peaks
  tells us that atoms are distributed among all nine levels resulting
  in a low orientation $p=0.346$. The line width 9.4Hz is a direct
  measure of the decay rate of spin coherence. According to
  Eq.~(\ref{eq:g_F}) the corresponding $F=3$ signal is approximately
  1kHz away and does not interfere here. The inset shows the level
  structure of the ground states of cesium following
  Eq.~(\ref{eq:exact_energy}). Our experiment is carried out around
  $10^{-4}$T which is far into the linear regime.  However, with a
  sufficiently good resolution the quadratic effect is visible.}
\label{fig:intro_qz}
\end{figure}

\subsection{Spin State Evolution in the Process of the 
            Double Magnetic Resonance}
\label{sec:descr-spin-state}
Let us consider an ensemble of atoms in a Zeeman multiplet in the
ground state of an alkali, in our case this is the $F=4$ state in
cesium. Applying an external magnetic field, the energy of each
magnetic sublevel changes as shown in the inset of
Fig.~\ref{fig:intro_qz}. For low fields the splitting between two
levels grows linearly with field strength with small quadratic
corrections given by $\nu_{\mathrm{QZ}} =
2\nu_{\mathrm{L}}^2/\nu_{\mathrm{hfs}}$, where $\nu_{\mathrm{L}}$ is
the Zeeman splitting, or the Larmor frequency, and
$\nu_{\mathrm{hfs}}$ is the hyperfine splitting. This is known as the
quadratic Zeeman effect, for completeness we have summarized the most
important results in appendix~\ref{sec:quadr-zeem-effect}.

Applying RF-magnetic fields we can induce transitions between magnetic
sublevels. Our spectral resolution is sufficiently high in order to
observe the small energy shifts caused by the quadratic Zeeman effect,
see Fig.~\ref{fig:intro_qz}.  We see eight distinct lines
corresponding to transitions between two adjacent levels among the
$2F+1$ possible levels in the $F=4$ state. The connection between the
spin state and the spectrum in Fig.~\ref{fig:intro_qz} will be derived
in the following, we will be able to fully characterize the spin state
in many situations important for other experiments.

We characterize the state of atoms by the density operator
$\densop{ij}$ given by
\begin{equation}
\densop{ij} = \frac{1}{N}\sum_{k=1}^N \densop{ij}^{(k)} 
            = \frac{1}{N}\sum_{k=1}^N \ket{i}_k\bra{j}_k 
\end{equation}
where $i,j= -F,-F+1,\ldots,F$ parametrizes the magnetic sublevels
$\ket{i}$ or $\ket{j}$ in Dirac notation, and the sum is done over all
individual atoms. With the $z$-axis as quantization axis we may
express the total macroscopic angular momentum of the atoms in the
hyperfine state $F$ as
\begin{subequations}
\begin{align}
\notag
\F{x} &= \frac{1}{2} \left\{\F{+} + \F{-}\right\} = \\
      &= N\sum_{m=-F}^{F-1} \frac{C(F,m)}{2}    \label{Fx}
         \left\{\densop{m+1,m}+\densop{m,m+1}\right\} \\
\notag
\F{y} &= \frac{1}{2i}\left(\F{+} - \F{-}\right) =  \\
      &= N\sum_{m=-F}^{F-1} \frac{C(F,m)}{2i}   \label{Fy}
         \left\{\densop{m+1,m}-\densop{m,m+1}\right\} \\
\F{z} &= N\sum_m m\densop{mm}  \label{Fz}
\end{align}
\end{subequations}
where $C(F,m) = \sqrt{F(F+1)-m(m+1)}$ and $\F{\pm}$ are
raising/lowering operators for the spin along $z$. Note, if we are
interested in calculating the spin components of $\vec{F}$, it
suffices to calculate the diagonal terms and the first off-diagonal
terms in the density matrix. This enables us to describe effects like
the \emph{orientation} of the spin state, see
Sec.~\ref{sec:modelling-spin-state}. In general, to describe aspects
of \emph{alignment} we also need second off-diagonal terms. We
restrict our description to diagonal and first off-diagonal terms in
the following.

The Hamiltonian of spins subject to a magnetic field $\vec{B}$ can be
written
\begin{equation}
\H = g_F \mu_B \vec{F}\cdot\vec{B} + O(B^2)
\end{equation}
where the second order correction is described in
appendix~\ref{sec:quadr-zeem-effect}. We will partly include this second
order correction, since it will prove to be important for the energy
levels. We will place a constant bias magnetic field with strength
$B_{\text{bias}}$ along the $z$-axis and apply an RF-magnetic field
$|B_{RF}|\cos(\omega t+\phi)$ along the $x$-axis. The Hamiltonian may
now be written (for a single atom)
\begin{equation}
\begin{split}
\H &= \sum_{m=-F}^F \hbar\omega_m\cdot\densop{mm}  \\
&+\frac{g_F\mu_B}{4N}\left(\F{+}B_{RF}\expml + \F{-}B_{RF}^*\exppl\right)
\end{split}
\end{equation}
where $B_{RF} = |B_{RF}|e^{-i\phi}$ is the complex amplitude. The
first term is primarily $F_z B_z$, but we take the second order
corrections into account by explicitly stating the energy levels
$\hbar\omega_m$ of the $m$'th sublevel. The second term is $\F{x} B_x$
originating from the $x$-polarized RF-magnetic field which induces
transitions between the magnetic sublevels. The rotating wave
approximation has been made here. We may wish to write the Hamiltonian
entirely in terms of the density operators $\densop{ij}$:
\begin{equation}
\label{eq:Hamiltonian}
\begin{split}
\H &= \sum_{m=-F}^F \hbar\omega_m\cdot\densop{mm}  \\
&+ \frac{g_F\mu_B}{4}\sum_{m=-F}^{F-1}
\left(C(F,m)\densop{m+1,m}B_{RF}\expml + \text{h.c.}\right)
\end{split}
\end{equation}
The equations of motion are now determined by
\begin{equation}
\label{eq:Equation_of_motion}
\frac{\partial \densop{ij}}{\partial t}  
= \frac{1}{i\hbar}\commutator{\densop{ij}}{\H} + \text{ decay terms}
\end{equation}
where the first term is the coherent evolution of the system, and the
interaction with the environment will be put in by hand as decay
terms.
\subsection{Solution of equations of motion}
\label{sec:solut-equat-moti}
We will now solve equations~(\ref{eq:Hamiltonian})
and~(\ref{eq:Equation_of_motion}), and to illuminate the method for
solving these equations, we will pick out a single example and work it
out thoroughly. The time derivative of e.g.~$\densop{12}$ is
\begin{equation}
\label{eq:solve_dens_eq_1}
\begin{split}
\frac{\partial \densop{12}}{\partial t}  
&= \frac{1}{i\hbar}\commutator{\densop{12}}{\H}
 - \Gamma/2\cdot\densop{12} \\
&= -i(\omega_2-\omega_1)\densop{12} - \Gamma/2\cdot\densop{12} \\ 
&\quad + \frac{i g_F \mu_B}{4\hbar}\left\{C(F,1)
[\densop{22}-\densop{11}] B_{RF}e^{-i\omega t} \right.\\
&\quad + \left.[C(F,0)\densop{02}
 - C(F,2)\densop{13}]B_{RF}^* e^{i\omega t}\right\}
\end{split}
\end{equation}
where we have just inserted the Hamiltonian~(\ref{eq:Hamiltonian})
into~(\ref{eq:Equation_of_motion}) and added the decay term,
$-\Gamma/2\cdot\densop{12}$. We will restrict ourselves to a
description of spins in the case where $F_x, F_y \ll F_z$, i.e.~the
angle $\theta$ that the spins deviate from being oriented along the
$z$-axis is much less than unity. From equations~(\ref{Fx}-\ref{Fz})
this can roughly be written as $\densop{m+1,m} \approx \theta\cdot
\densop{m,m}$, and following the same lines $\densop{m+2,m} \approx
\theta^2\cdot \densop{m,m}$. It is then justified to neglect the
coherences $\densop{02}$ and $\densop{13}$ in the above equation. For
brevity we will define $\omega_{21} = \omega_2-\omega_1$, which is the
frequency corresponding to the transition from $m_F=2$ to $m_F=1$.
This frequency is the Larmor frequency and it is fast compared to the
inverse time scale for dynamical evolution of the spin state. Since
the RF-magnetic field frequency $\omega$ will be in the vicinity of
$\omega_{21}$ it will be convenient to define the slowly varying
operators
\begin{equation}
\label{eq:slow_operators}
\densop{ij} = \slowdensop{ij} e^{-i\omega t}
\end{equation}
Using this definition, equation~(\ref{eq:solve_dens_eq_1}) will turn into
\begin{equation}
\begin{split}
\label{eq:time_deriv_dens12}
\frac{\partial \slowdensop{12}}{\partial t}  
&= (i[\omega-\omega_{21}] - \Gamma/2)\slowdensop{12} \\
&+ \frac{i g_F \mu_B}{4\hbar}C(F,1)B_{RF}[\densop{22}-\densop{11}] 
\end{split}
\end{equation}
The constant $\Gamma$ will describe the decay of the \emph{transverse}
spin components. We will assume the time scale $T_2$ for the
transverse spin decay is much shorter than the time scale $T_1$ for
the longitudinal spin component (along the $z$-axis). Experimentally
we typically find $T_1 \approx$ 200-300ms and $T_2 \le 40$ms and the
approximation is justified. Then the operator $\slowdensop{12}$ will
follow $(\densop{22}-\densop{11})$ adiabatically which is expressed
mathematically by setting $\partial\slowdensop{12}/\partial t = 0$ in
the above equation. Alternatively, if constant pumping maintains $F_z$
the small angle condition $\theta \ll 1$ will ensure constant
$(\densop{22}-\densop{11})$. Then the above is simply the steady state
condition where transients have been damped away. The result for
$\densop{12}$ is then
\begin{equation}
\label{eq:solution_densop12}
\densop{12} = -\frac{i g_F \mu_B B_{RF} C(F,1) e^{-i\omega t}}
{4\hbar\cdot(i[\omega-\omega_{21}] - \Gamma/2)}[\densop{22}-\densop{11}]
\end{equation}
This method applies to all density operators $\densop{m,m+1}$, and
substituting into equations~(\ref{Fx}) and~(\ref{Fy}) we get
\begin{widetext}
\begin{subequations}
\begin{align}
\label{eq:Fx_solved}
\F{x} &= \Re \left\{ \frac{i g_F \mu_B B_{RF}N}{4\hbar} \sum_{m =
-F}^{F-1} \frac{[F(F+1)-m(m+1)]\cdot e^{i\omega t}}
{i(\omega_{m+1,m}-\omega)-\Gamma_{m+1,m}/2}
[\densop{m+1,m+1}-\densop{m,m}] \right\} \\ 
\label{eq:Fy_solved}
\F{y} &= \Im \left\{
\frac{i g_F \mu_B B_{RF}N}{4\hbar} \sum_{m = -F}^{F-1}
\frac{[F(F+1)-m(m+1)]\cdot e^{i\omega t}}
{i(\omega_{m+1,m}-\omega)-\Gamma_{m+1,m}/2} [\densop{m+1,m+1}-\densop{m,m}]
\right\}
\end{align}
\end{subequations}
\end{widetext}
These equations can be interpreted as a number $(2F)$ of two-level
systems that all interact with the driving RF-magnetic field. Two
adjacent magnetic sublevels $m$ and $m+1$ act as one two-level atom
with the usual Lorentzian response (resonance frequency
$\omega_{m+1,m}$ and line width $\Gamma_{m+1,m}$ FWHM). Each two-level
system does not respond with exactly the same weight which is
reflected in the factor $F(F+1)-m(m+1)$.  All the resonances add up
coherently to give the overall response of the spin state to the
RF-magnetic field. Note, that $\F{x}$ and $\F{y}$ oscillate at the
driving frequency $\omega$ and not the ``natural'' frequencies
$\omega_{m+1,m}$. This is a steady state behavior, just like the
forced harmonic oscillator when transients have been damped. In
section \ref{pulsed_exp} we will comment on non-steady state behavior
of the spin system.

\subsection{Modeling the spin state}
\label{sec:modelling-spin-state}
\begin{figure*}
\includegraphics{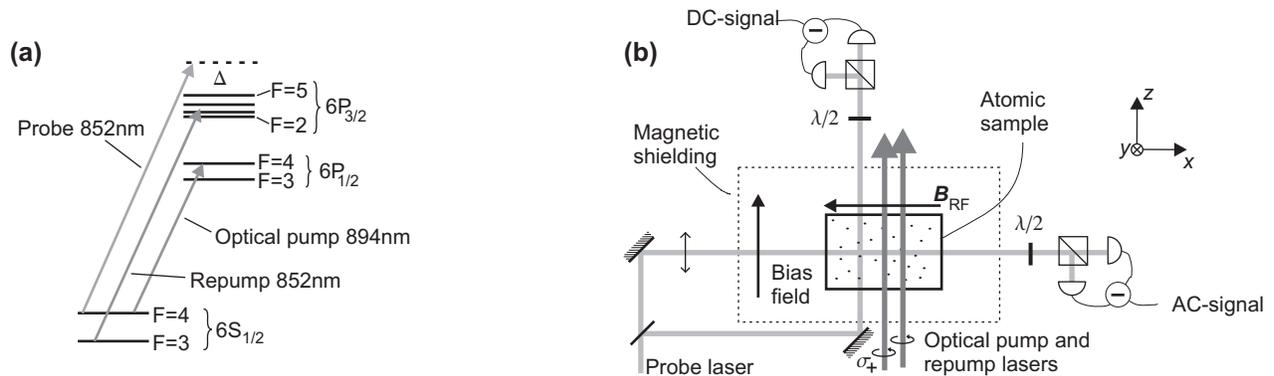}
\caption{{\bf (a)} The lasers used in the experiment together with the
level scheme of cesium. The $F=4$ ground state is with its nine
sublevels of main interest in this experiment. {\bf (b)} The
experimental setup. The atomic ensemble is illuminated with optical
pumping and repumping lasers with $\sigma_+$-polarized light parallel
to a static bias magnetic field. This orients the spins along the
$z$-axis and gives rise to a 325kHz Larmor precession. An RF-magnetic
field perpendicular to the spins induce small spin components in the
plane orthogonal to $z$ which are read out by polarization rotation of
a linearly polarized probe laser beam (giving an AC-signal). The spin
along $z$ can also be read out on a probe laser beam (giving a
DC-Faraday signal).}
\label{fig:exp_setup}
\end{figure*}
In the previous subsection we derived how the spin $\vec{F}$ responds
to an external RF-magnetic field $\vec{B}_{\mathrm{RF}}$. Our main
motivation is to use this knowledge to characterize the spin state,
i.e.~we wish to apply the field $\vec{B}_{\mathrm{RF}}$ and measure
the response $\F{x}$ or $\F{y}$ in order to deduce parameters like
$\densop{m,m}$, $\Gamma_{m+1,m}$ and so on. Now, for cesium in the
$F=4$ hyperfine ground state there are nine populations $\densop{m,m}$
and eight line widths $\Gamma_{m+1,m}$ together with the resonance
frequencies. To fit an experimentally measured spectrum (see
e.g.~Fig.~\ref{fig:intro_qz}) to all these parameters will be very
hard and in the following we will develop a model to significantly
reduce the number of free parameters. We will just tailor a model and
the justification for this model will be an experimental test.

Let us consider a case where we wish to orient all atomic spins
along the $z$-direction, i.e.~attempt to put many atoms into the $m=F$
substate. This can be done experimentally by illuminating the atoms
with circularly polarized light, as will be described in the
experimental section. It is then convenient to define the orientation
$p$ as an order of merit

\begin{equation}
p = \frac{1}{F} \sum_{-F}^F m\cdot\densop{m,m} = \frac{\F{z}}{NF}
\end{equation}
Note, that with this definition $p=1$ if all atoms are in the extreme
$m=F$ sublevel, and $p=0$ for a completely unpolarized sample with
$\densop{m,m} = 1/(2F+1)$ for all $m$. We try to let this be the only
parameter describing the relationship between the nine populations
$\densop{m,m}$. With the condition $\sum \densop{m,m} = 1$ we have
thus reduced eight free parameters to a single one.

Now, we describe ensembles of atoms and given $p$ we will assume that
the spin state maximizes the entropy $S =
-\mathrm{Tr}(\densop{}\ln\densop{})$. To find the individual
$\densop{m,m}$ we use the method of Lagrange multipliers. We must solve
\begin{gather}
\frac{\partial}{\partial\densop{m,m}}\left(S - \alpha\sum\densop{m,m}
- \beta\sum m\cdot\densop{m,m}\right)=0 \notag \\
\Rightarrow\quad 
\densop{m,m} = e^{-1-\alpha}\cdot e^{-\beta\cdot m}
\end{gather}
We now just need to adjust $\alpha$ and $\beta$ in order that
$\mathrm{Tr}(\densop{})=1$ and $p$ is as desired. Doing this is more
or less a computational problem and in principle not difficult. For
the physical understanding we just need to remember that we can write
$\densop{m,m} \propto \epsilon^m$ where $\epsilon = e^{-\beta}$ is a
parameter which is a function of $p$. This can go directly into
equations~(\ref{eq:Fx_solved}) and~(\ref{eq:Fy_solved}). We stress
that the above considerations are ment as an intuitive guideline to
understand what is really just an experimentally justified model. This
model is only valid in some specific cases, our example with circular
polarized pumping light. Our experience tells us that the model works
fine for large longitudinal relaxation times $T_1$ and not too strong
influence by the probe laser.

For the eight line widths $\Gamma_{m+1,m}$ in the case of cesium we
will make a model with two free parameters. First, a common line width
$\Gamma_{\text{com}}$ is assigned to all transitions independent of
$m$. The physical cause for this type of decay could be magnetic field
inhomogeneities, collisions, and loss mechanisms common to all
atoms. In addition, if we wish to create a well oriented sample with
$m$ approaching $F$ we will need to illuminate the atoms with resonant
circularly polarized light. In our case this is the 894nm $6S_{1/2},
F=4$ to $6P_{1/2}, F'=4$ line. This light causes excitations from the
atomic ground sublevel $m$ with a rate $\gamma_m\propto
|\left<4,m,1,1|4,m+1\right>|^2 = (4-m)(5+m)/40$, where the second term
is the square of Clebsch-Gordan coefficients. For a magnetic
transition between ground sublevels $m$ and $m+1$ the resonant pumping
light will contribute to the line broadening proportional to $\gamma_m
+ \gamma_{m+1}$. Thus we define the width $\Gamma_{\text{pump}}$
caused by the optical pumping process such that
\begin{equation}
\label{eq:def_gamma_com_pump}
\Gamma_{m+1,m} = \Gamma_{\text{com}} + \Gamma_{\text{pump}}\frac{19-2m-m^2}{4}
\end{equation}
where the normalization is such that for the $m=3 \rightarrow m=4$
transition we have $\Gamma_{4,3} = \Gamma_{\text{com}} +
\Gamma_{\text{pump}}$.

Finally, we must have the resonant frequencies as parameters in our
model. We will write this as a central frequency
$\omega_{\text{center}}$ and a splitting $\omega_{\text{split}}$ such
that
\begin{equation}
\label{eq:freq_vs_m}
\omega_{m+1,m} = \omega_{\text{center}} 
+ \omega_{\text{split}}\left(m+\frac{1}{2}\right)
\end{equation}
Theoretically we should have $\omega_{\text{split}} =
2\pi\cdot\nu_{\text{QZ}}$ (see equation~(\ref{def:QZ_splitting})) but
we keep it as a free parameter since in practice this splitting may
differ slightly from the theoretical value because of Stark
splittings.

To sum up, a possible description of the ground spin state involves
the total spin $\F{z}$ and the orientation $p$ together with the line
widths $\Gamma_{\text{com}}$ and $\Gamma_{\text{pump}}$, and the
frequencies $\omega_{\text{center}}$ and $\omega_{\text{split}}$. An
equivalent but computationally easier way to represent $\F{z}$ and $p$
is to use the number of atoms $N_4 = N\densop{44}$ of atoms in $m=4$ as one
parameter and the parameter $\epsilon$ such that the population
$N_m$ can be expressed as $N_m = N\densop{m,m} =
N_4\epsilon^{(4-m)}$.
\section{Experiment}
\label{sec:experiment}
In this experimental section we will first describe the experimental
setup in order to understand what is actually measured and how the
measurements relate to the equations of
section~\ref{sec:theor-descr}. Having this in place we comment on the
requirements that actually allow us to resolve the individual
sublevels as seen in Fig.~\ref{fig:intro_qz}.
\subsection{Experimental setup}
\label{sec:experimental-setup}
The experimental setup and the laser level scheme is shown in
Fig.~\ref{fig:exp_setup}. Referring to part {\bf (a)} of the figure,
we have a Ti:Sapphire laser probe detuned by $\Delta \approx 1$GHz
from the $6S_{1/2},F=4 \rightarrow 6P_{3/2},F=5$ transition at 852nm.
This laser is sensitive to the state of atoms in the $F=4$ ground
state and will be used to probe these atoms. To control the state of
atoms in the $F=4$ ground state we have two home built grating
stabilized diode lasers. The \emph{repump laser} is tuned to the
$6S_{1/2},F=3 \rightarrow 6P_{3/2},F=4$ transition at 852nm. This
mainly serves to remove atoms from the otherwise dark $F=3$ ground
state. The \emph{optical pump laser} tuned to the $6S_{1/2},F=4
\rightarrow 6P_{1/2},F=4$ transition at 894nm plays the most important
role in distributing atoms among the sublevels of the $F=4$ ground
state. Note, that atoms in the extreme $F=4$, $m_F=4$ state will not
absorb light from any of the two lasers.

Now turning to Fig.~\ref{fig:exp_setup}b, we place a paraffin coated
vapor cell containing cesium inside a shield of $\mu$-metal to isolate
the atoms from laboratory and Earth magnetic fields. A static magnetic
field of strength approximately 0.9 Gauss is then applied in the
$z$-direction and the repump and optical pump lasers are illuminating
atoms along that direction. The polarization of these lasers can be
adjusted, and by choosing circular polarization we can orient the
atomic spins along the $z$-direction.

The probe laser is split into two beams. One of them is directed along
the $x$-direction transverse to the atomic magnetization created by the
repump and optical pump lasers. This probe laser is linearly polarized
along the $z$-direction and will undergo polarization rotation
proportional to the spin $\F{x}$ along the $x$-axis (see e.g.
\cite{happer:67,julsgaard:03}). By suitable arrangement of a
$\lambda/2$-plate and a PBS, the difference signal of two photo
detectors will be proportional to the polarization rotation (for small
angles).

The atomic spins $\F{x}$ along the $x$-axis will have zero mean value
unless an RF-magnetic field is applied transversely to the static
magnetic field. We apply such a field at frequency $\omega$ (we shall
call this frequency the local oscillator frequency), and the motion of
$\F{x}$ will now be described by equation~(\ref{eq:Fx_solved}).  We
may write the outcoming AC-signal as $i(t) = \alpha\cdot \F{x} =
\alpha\cdot \mathrm{Re}\{A(t)\}$ where $\alpha$ is a constant
depending on beam geometry, laser detuning and intensity
\cite{julsgaard:03}, and $A(t)$ reflects the curly bracket of
equation~(\ref{eq:Fx_solved}). We know from this equation that $A(t)
\equiv A(\omega) e^{i\omega t}$ will posses only a single frequency
component, namely the local oscillator frequency $\omega$ driving the
transverse spins $\F{x}$ and $\F{y}$ away from zero. The amplitude of
this frequency component is experimentally measured by inserting the
photo current $i(t)$ into a lock-in amplifier and decomposing the
signal into sine and cosine components:
\begin{gather}
\notag
i(t) = \alpha\cdot\Re\{A(\omega)e^{i\omega t} \} \\
= \alpha\cdot\left(
\Re\{A(\omega)\}\cos(\omega t) - \Im\{A(\omega)\}\sin(\omega t)\right)
\end{gather}
The lock-amplifier may give the sum of the squared amplitudes of the
sine and cosine components which in our case will be exactly
$\alpha^2|A(\omega)|^2$. We shall call this signal our
\emph{magneto-optical resonance signal} at frequency $\omega$
(MORS($\omega$) in short). Combining the above with
equation~(\ref{eq:Fx_solved}) and ignoring irrelevant constants we
find
\begin{widetext}
\begin{equation}
\label{def:MORS}
\text{MORS}(\omega) = \text{const}\cdot \left|
N \sum_{m=-F}^{F-1} \frac{[F(F+1)-m(m+1)]}
{i(\omega_{m+1,m}-\omega)-\Gamma_{m+1,m}/2}
[\densop{m+1,m+1}-\densop{m,m}] \right|^2
\end{equation}
\end{widetext}
The second part of the probe laser beam in Fig.~\ref{fig:exp_setup}b
is linearly polarized and is directed along the static magnetic field
direction and will be subject to polarization rotation proportional to
the atomic spin $\F{z}$. The angle of rotation $\theta_{\mathrm{DC}}$
(denoted the DC-Faraday rotation angle) can be measured and we have in
analogy with~(\ref{def:MORS})
\begin{equation}
\label{eq:DC-signal}
\theta_{\mathrm{DC}} = \text{const}\cdot \F{z} =
\text{const}\cdot N \sum_{m=-F}^F m \densop{m,m}
\end{equation}
With the magneto-optical resonance signal, the DC-Faraday rotation
angle, and the methods of section~\ref{sec:modelling-spin-state} at
hand we will be able to say much about the number of atoms $N$ of the
sample, the populations $\densop{m,m}$ and the decay rates
$\Gamma_{m+1,m}$ of the coherences $\densop{m,m+1}$.
\subsection{Resolving the different Zeeman lines}
\label{sec:resolv-diff-zeem}
To resolve the different Zeeman lines we obviously need
$\Gamma_{m+1,m} \lesssim \omega_{\mathrm{QZ}}$. There are many
contributions to the decay of the transverse coherences, collisions
between atoms, collisions with the cell walls, power broadening by
laser light, dephasing by magnetic field inhomogeneities.

Atoms are kept in a paraffin coated glass cell which prevents atoms
from depolarizing when hitting the walls. This method can increase the
coherence time up to roughly 1 second \cite{bouchiat:66}. However, our
record is somewhat lower at 50ms.

The probe laser beams give only a marginal contribution to the decay
if they are detuned sufficiently far from the resonance or has a low
intensity. We estimate the rate $\Gamma_{\mathrm{ph}}$ at which an
atom scatters photons by the two-level atom result (see
e.g.~\cite{metcalf})
\begin{equation}
\Gamma_{\mathrm{ph}} = \frac{\gamma}{2}\frac{s}{1+s}
\approx 
\frac{3I\lambda^3\gamma^2}{16\pi^2\hbar c\Delta^2}
\end{equation}
where $s = \frac{I/I_{\mathrm{sat}}}{1+(2\Delta/\gamma)^2}$ is the
saturation parameter. $I$ is the beam intensity with $I_{\mathrm{sat}}
=2\pi^2\hbar c \gamma/3\lambda^3$ being the saturation intensity.
$\lambda$ is the optical wave length, $\gamma$ is the natural line
width of the optical transition, and $\Delta$ is the detuning (assumed
much greater than $\gamma$ in the last step of the equation).
In our case we operate at up to 1mW/cm$^2$ and the detuning is
typically around 1GHz which contributes the order of
$\Gamma_{\mathrm{ph}}= 100$Hz to the decay rate according to the above
estimate. We usually observe a somewhat smaller width which probably
can be attributed to the crude two-level atom approximation.
Nonetheless, to be able to see the magneto-optical resonance signal a
considerably weaker probe can be used and the power broadening can
easily be reduced below 1Hz.

The repump laser beam is detuned roughly 9GHz from the $F=4$ ground
state and thus plays a very minor role even at strong intensities. The
optical pump laser is directly on resonance with the $F=4$ ground
state sublevels and then only a moderate intensity can be allowed if
different Zeeman lines should be resolved. The quantitative aspects of
the optical pump laser power broadening was discussed in
equation~(\ref{eq:def_gamma_com_pump}).

\begin{figure}
\psfrag{taga}{$\Gamma_{\mathrm{inh}}[\mathrm{Hz}] = 8.7(3) + 0.0158(2)
               \left(\frac{\partial B}{\partial z}
               \left[\frac{\mathrm{mG}}{\mathrm{m}}\right]\right)^2$}
\psfrag{tagb}{$B_0 = 0.93$G}
\includegraphics{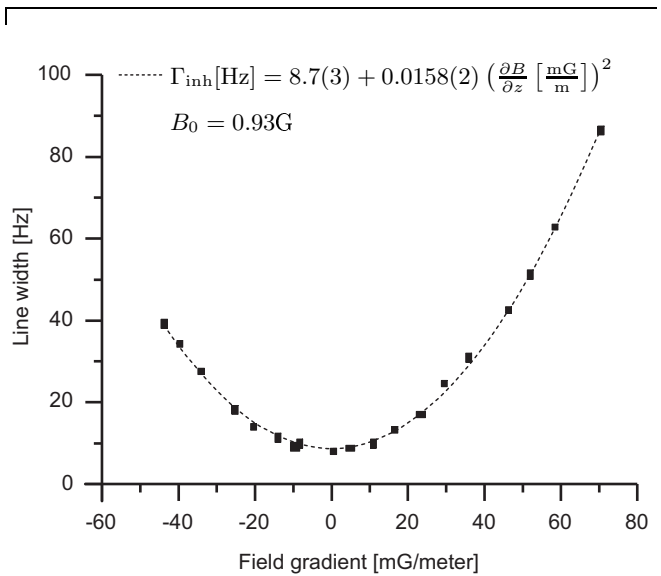}
\caption{The measured line width as a function of an applied magnetic
field gradient. We confirm the scaling derived in
equation~(\ref{eq:theory_inh_width}). The 8.7Hz minimum is set by
collisions, power broadening or possible residual inhomogeneities and
the additional quadratic part stems from the applied magnetic field
inhomogeneity.}
\label{fig:inhom_mag_field}
\end{figure}

The magnetic field must be homogeneous to a high degree. Since atoms
are moving around in the magnetic field, different atoms will
experience different magnetic field strengths and hence different
Larmor frequencies. To estimate the role of a possible gradient
$\partial B/\partial z$ we use the following simple model (see
e.g.~\cite{kleppner:62}).
First, divide the atomic sample into two parts, 1 and 2, along the
bias magnetic field direction. If the sample length is $L$ and the
bias field has strength $B_0$, the field strength in the two parts
will be of order $B_0 \pm \partial B/\partial z \cdot L$ and the
difference in Larmor frequency will be $(g_F\mu_B/\hbar)\partial
B/\partial z\cdot L$ according to~(\ref{eq:E_zeeman_1st_order}). We
follow an atom during the time $T$ it takes for it to decohere. If $v$
is a typical speed of the atomic motion, the number of visits $n_1$ in
part 1 or $n_2$ in part 2 will be of order $Tv/L$, since each visit
has duration $L/v$. The difference $n_1-n_2$ has mean zero and standard
deviation of the order std$(n_1-n_2) =\sqrt{Tv/L}$. Thus the
uncertainty $\delta \phi$ in the accumulated phase during Larmor
precession is
\begin{align}
\notag
\delta\phi &\approx\frac{g_F \mu_B}{\hbar}\cdot\frac{\partial B}{\partial z}L 
              \cdot \frac{L}{v}\cdot \mathrm{std}(n_1-n_2) \approx 1 \\
\Rightarrow \quad \Gamma_{\mathrm{inh}} &\approx \frac{1}{T} \approx
\left(\frac{g_F \mu_B}{\hbar}\right)^2\frac{L^3}{v}
\left(\frac{\partial B}{\partial z}\right)^2
\label{eq:theory_inh_width}
\end{align}
In the first line we set $\delta\phi$ equal to unity since this is the
situation after the time of decoherence $T$. We see that the
broadening $\Gamma_{\mathrm{inh}}$ by inhomogeneities scales
quadratically with the field gradient. If we take $g_F \approx 1/4$
(see eq.~(\ref{eq:g_F})), $L = 0.030$m, $v = \sqrt{k_B
T/m_{\mathrm{Cs}}} = 137$m/s at $T = 300$K, we get $g_F\mu_B/\hbar =
350$Hz/mG and expect the broadening to be $\Gamma_{\mathrm{inh}} =
0.024 \mathrm{Hz}\cdot\mathrm{m}^2/\mathrm{mG}^2 \cdot (\partial
B/\partial z)^2$.

The experimental investigation can be seen in
Fig.~\ref{fig:inhom_mag_field} and we definitely confirm the scaling
law predicted above. The numbers match within a factor of two which
puts some confidence to our simple model but this is probably also
partly luck since we were very crude in the model w.r.t. factors of 2
and $\pi$. Comparing the experimental result with the splitting due to
the quadratic Zeeman effect~(\ref{def:QZ_splitting}) we find for our
particular setup that in order to have $\Gamma_{\mathrm{inh}} <
\nu_{\mathrm{QZ}}$ we must have $1/B\cdot\partial B/\partial z\cdot L
< 1.2\cdot 10^{-3}$.

\subsection{Confirming the spin modeling}
\label{sec:conf-spin-model}
\begin{figure*}
  \psfrag{taga}{$\begin{matrix}
                 \Gamma_{\mathrm{com}} = 9.4\mathrm{Hz} \\
                 \Gamma_{\mathrm{pump}} = 0.0\mathrm{Hz} \\
                 J_z = 0.248\mathrm{[a.u.]} \\
                 p = 0.823
                 \end{matrix}$} 
  \psfrag{tagb}{$\begin{matrix}
                 \Gamma_{\mathrm{com}} = 9.4\mathrm{Hz} \\
                 \Gamma_{\mathrm{pump}} = 5.5\mathrm{Hz} \\
                 J_z = 0.336\mathrm{[a.u.]} \\
                 p = 0.967
                 \end{matrix}$} 
  \includegraphics{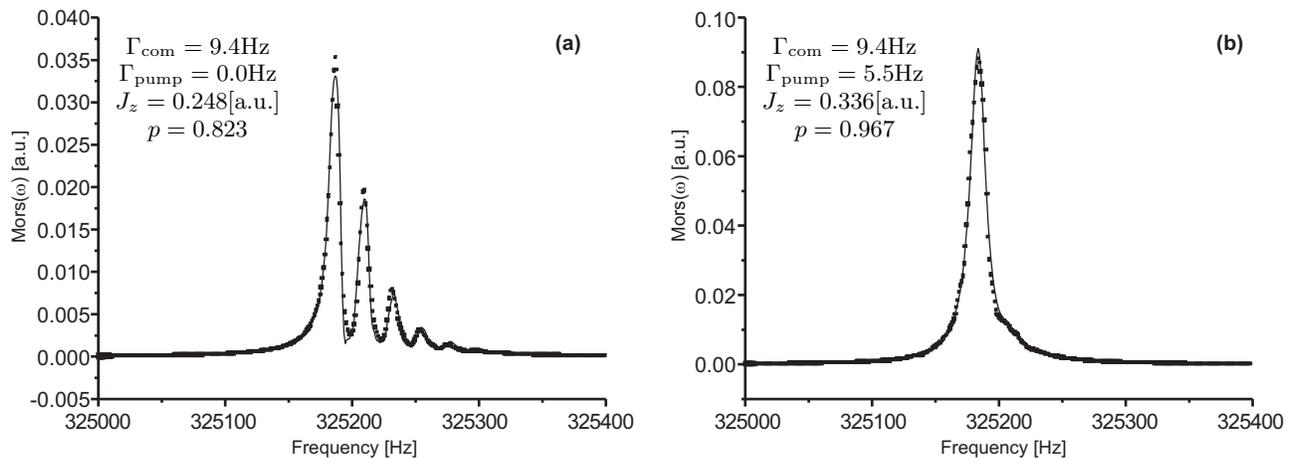}
\caption{Two examples of experimental (dots) and fitted (solid line) traces.
  The left graph was obtained with pure $\sigma$-polarized repump
  laser and no optical pumping. On the right graph a small amount of
  optical pumping light is added giving rise to a non-zero
  $\Gamma_{\mathrm{pump}}$. Here one can, on careful inspection,
  observe that the two leftmost peaks have different widths. Note
  also, that the height has grown by a factor of three compared to the
  graph on the left.}
\label{fig:mag_traces}
\end{figure*}
We will now give experimental support to the theoretical derivations
and the modeling of the spin state as described in
section~\ref{sec:theor-descr}. We first focus on the validity of the
simple model introduced in subsection~\ref{sec:modelling-spin-state},
i.e.~we test the dependence $\densop{m,m} \propto \epsilon^m$, the model
for the line widths~(\ref{eq:def_gamma_com_pump}), and the frequency
distribution~(\ref{eq:freq_vs_m}).

With the setting as in Fig.~\ref{fig:exp_setup}b we leave the optical
pump laser off and vary the polarization of the repump laser. This
will create different distributions of $\densop{m,m}$ and we record
the MORS. In Fig.~\ref{fig:intro_qz} we see an example where all
eight lines are clearly visible, the polarization of the repump laser
was here quite far away from being purely circular. The dots are
experimental points and the solid line is a fit to the
model~(\ref{def:MORS}) with $N$, $\epsilon$, $\Gamma_{\mathrm{com}}$,
$\omega_{\mathrm{center}}$, and $\omega_{\mathrm{split}}$ as free
parameters. $\Gamma_{\mathrm{pump}}$ is set to zero. We see that the
solid line matches the experimental points very well. Note, that
$\epsilon$ corresponding to $p = 0.346$ is the only parameter really
describing the relative strength of the individual peaks, while the
other parameters are common to all peaks. The center frequency
$\omega_{\mathrm{center}} \approx 325250$Hz is set by the magnetic
field (or to be true, this frequency was a quiet place in terms of
laser noise for other experiments and thus our detectors was tuned to
this frequency). The splitting $\omega_{\mathrm{split}} = 22$Hz is
close to the 23Hz expected from Eq.~(\ref{def:QZ_splitting}). The
small deviation is due to Stark shifts from the laser beams. Finally,
we find $\Gamma_{\mathrm{com}} = 9.4$Hz (FWHM). This corresponds to a
life time of the spin coherence of 34ms.

The next example is recorded with the repump laser purely circular and
optical pump laser still off. The spectrum can be seen in
Fig.~\ref{fig:mag_traces}a. Now the spectrum is displaced much more to
one side and the fit gives $p = 0.823$. This single parameter still
seems to describe the shape with good accuracy.

The third example we will show is seen in Fig.~\ref{fig:mag_traces}b.
Here the situation is as before but now with a weak optical pump
present. We observe an additional broadening of the left most peak by
$\Gamma_{\mathrm{pump}} = 5.5$Hz and we also note that the second peak
seems much broader (should have an additional broadening by 15.1Hz
according to Eq.~(\ref{eq:def_gamma_com_pump})). Since the fit and
the experimental points follow each other very well, we get support
for the modeling of $\Gamma_{\mathrm{pump}}$. The orientation $p =
0.967$ shows that we are very close to have all atoms in $m_F = 4$
with only a moderate amount of optical pumping light.

\begin{figure*}[t]
  \includegraphics{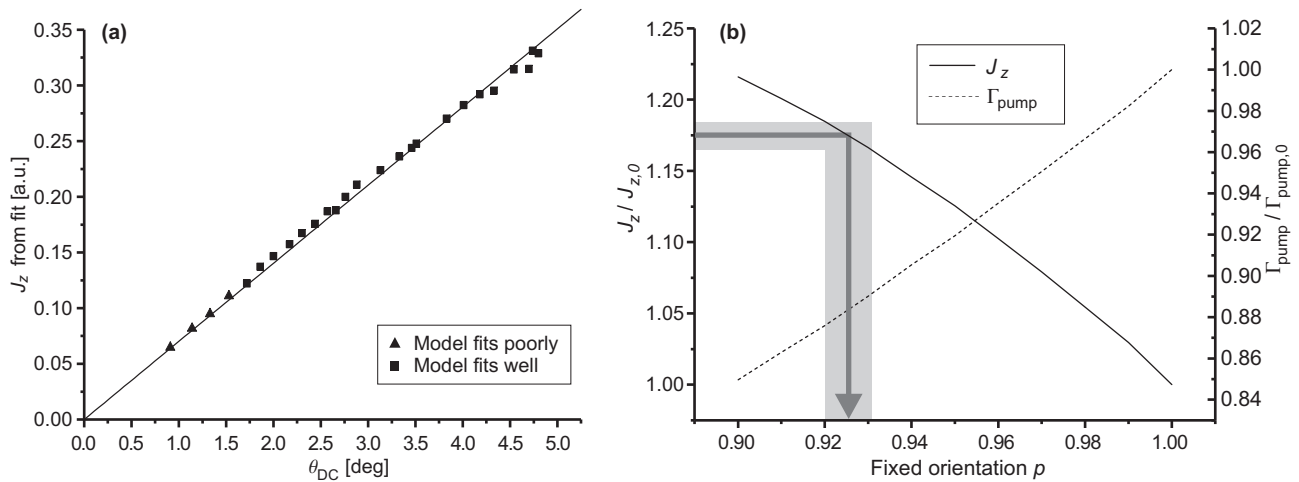}
\caption{\textbf{(a)} Consistency check of the models. Fits to different
  spectra give an estimate of $J_z$. This can be compared directly to
  the independently measured DC-Faraday rotation signal
  $\theta_{\mathrm{DC}}$ which is proportional to $J_z$. We indeed
  observe a straight line through the origin. Note, the model
  description $\densop{m,m}\propto\epsilon^m$ proved to be less
  accurate for the lowest four points, but by coincidence the points
  still fit well. \textbf{(b)} The interdependence of $p$, $J_z$ and
  $\Gamma_{\mathrm{pump}}$ in the limit where $\Gamma_{\mathrm{pump}}$
  dominates both the common width $\Gamma_{\mathrm{com}}$ and the
  quadratic Zeeman splitting $\omega_{\mathrm{split}}$. Different
  choices of fitting parameters will lead to a satisfactory fit and
  additional information is needed to place precise bonds on $p$. If
  for example $J_z$ is measured independently with an accuracy of
  $2\%$ the orientation can be defined within $1\%$ in the example
  shown.}
  \label{fig:check_model}
\end{figure*}
In the three examples described above we also get a fit for the number
of atoms $N$ on a relative scale (the constant in front of
Eq.~(\ref{def:MORS}) depends on many experimental parameters so we do
not wish to calculate it in absolute units). Multiplying this $N$ with
the fitted orientation $p$ gives the total spin $J_z$ (on a relative
scale). Now, the DC-Faraday rotation signal $\theta_{\mathrm{DC}}$
(see~(\ref{eq:DC-signal})) gives an independent measure of $J_z$ and
we may compare the fitted $J_z$ with $\theta_{\mathrm{DC}}$ to get
another consistency check on the model. This is shown in
Fig.~\ref{fig:check_model}a where we plot the fitted $J_z$ as a
function of $\theta_{\mathrm{DC}}$. The lowest points are taken with
the repump laser only and varying repump polarization. The upper eight
points are taken with purely circular optical pump of increasing
intensity in addition to a purely circular repump laser. We see a very
nice agreement between the fitted and the directly measured values
giving strong support to both the derivations leading to
Eq.~(\ref{def:MORS}) and the modeling of the spin state described in
subsection~\ref{sec:modelling-spin-state}.
\subsection{Unresolved lines}
\label{sec:unresolved}
The spectra shown so far have been more or less well resolved which
enabled us to directly determine the orientation $p$. Now, how much
information can we extract if the line widths are much broader than the
quadratic Zeeman splitting $\omega_{\mathrm{split}}$?

First, assume that all atoms are subject to decoherence with the same
rate described by $\Gamma_{\mathrm{com}} \gg \omega_{\mathrm{split}}$
and decay from pumping light is a small contribution
$\Gamma_{\mathrm{pump}} \approx 0$. Then Eq.~(\ref{def:MORS}) reduces to
\begin{equation}
  \label{eq:MORS_large_Gamma_com}
  \mathrm{MORS}(\omega) = \mathrm{const}\cdot\left|
2N\frac{\sum m\densop{m,m}}
{i(\omega_0 - \omega) + \Gamma_{\mathrm{com}}}\right|^2
\propto \left| J_z \right|^2
\end{equation}
We see that in this case the spectrum will be a single Lorentzian the
size of which is only depending on $J_z$. In this case the independent
measure from the DC-Faraday signal would only contribute exactly the
same information and we will not be able to deduce the orientation $p$.

On the other hand, if $\Gamma_{\mathrm{pump}}$ dominates
$\Gamma_{\mathrm{com}}$ and $\omega_{\mathrm{split}}$ we will get a
signal that depends on the internal atomic spin state. To examine this
approximation we set $\Gamma_{\mathrm{com}} = \omega_{\mathrm{split}}
= 0$ and try to fit the rest of the parameters to a spectrum which is
a perfect Lorentzian. The correct fitting parameters of course have
$p=1$ and $\Gamma_{\mathrm{pump}}$ equal to the Lorentzian width but
in practical life other sets of parameters will also fit the spectrum
to some extent. We find that orientations in the range $p = 0.9$ to
$p=1$ result in an agreement one would find reasonable if the spectrum
was an experimental trace. By fixing $p$ to a value in this range, the
values given from the fit of $J_z$ and $\Gamma_{\mathrm{opt}}$ are
shown in Fig.~\ref{fig:check_model}b. We see that if we can estimate
one of the parameters $J_z$ or $\Gamma_{\mathrm{opt}}$ independently
we should be able to calculate the orientation $p$. For instance, a
measurement of $J_z$ to an accuracy of 2\% will fix the orientation
$p$ to 1\%. One only needs to have one fix point, e.g.~if one knows
that we have $p=1$ perfectly in one case, or if one can reduce
$\Gamma_{\mathrm{pump}}$ to the point where the spectral lines become
resolved and a calibration like Fig.~\ref{fig:check_model} can be
performed. Experimentally we have seen orientations better than
$p=0.98$ for atomic densities around $10^{11}\mathrm{cm}^{-3}$. The
optical pumping laser at the 894nm $D1$-line is essential to this
achievement. We have tried to optically pump on the $D2$-line with
somewhat lower orientation as a result (a little above $p=0.9$). A
possible explanation is that the rescattered light on the $F=4$,
$m_F=4 \rightarrow F=5$, $m_F=5$ transition from one atom affects the
state of other atoms. Indeed, according to \cite{tupa:87}, even with a
dark state when using 894nm pumping light one would expect problems
with densities higher than $\rho_{\mathrm{C}} = (\sigma R)^{-1}$
because radiation will be trapped inside the sample. Here $\sigma$ is
the cross section for light absorption and $R$ is the extent of the
gas sample.  Our atomic sample is Doppler broadened with the width
$\delta\nu_{\mathrm{D}}= 378$MHz. With a natural line width of the
894nm $D1$-transition of $\gamma = 4.6$MHz and a sample extent of
$R=3$cm we estimate the critical density $\rho_{\mathrm{C}}$ to be
roughly $\rho_{\mathrm{C}}\approx
[\lambda^2/2\pi\cdot\gamma/\delta\nu_{\mathrm{D}}\cdot R]^{-1} =
2\cdot 10^{11}\mathrm{cm}^{-3}$. We see that we are in the regime
where radiation trapping may be a limiting factor, but the experiments
tell us that the limitations are still small.

\section{Pulsed experiments}
\label{pulsed_exp}
\begin{figure}[t]
  \psfrag{tag}{$\begin{matrix}
                 \Gamma_{\mathrm{pump}} = 770\mathrm{Hz} \\
                 \Gamma_{\mathrm{dark}} = 18\mathrm{Hz} \\
                 \Gamma_{\mathrm{probe}} = 2\mathrm{Hz} \\
                 \end{matrix}$} 
  \includegraphics{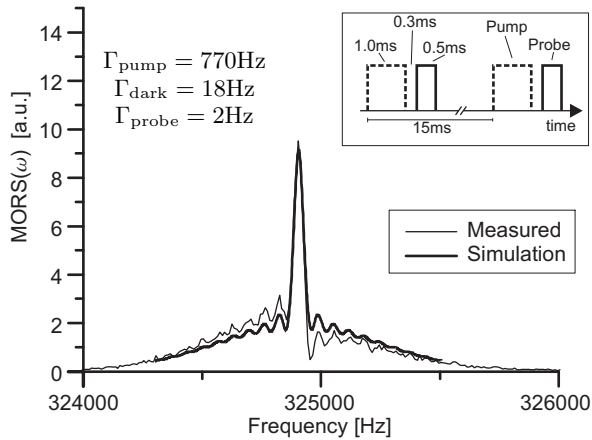}
  \caption{An example of the magneto-optical resonance signal in a pulsed 
           regime. The spin state is close to the maximal state with 
           $m_F = 4$ as assumed in the text. Note the ripples which
           have frequency spacing $(15\mathrm{ms})^{-1}$. The decay rate
           $\Gamma$ takes different values during one measurement cycle
           depending on which lasers are on. The timing of laser pulses is
           shown in the inset and the corresponding decay rates shown on
           the left part of the figure.}
  \label{fig:pulsed_graph}
\end{figure}
All previous derivations and measurements have been carried out in cw
settings, i.e.~Eqs.~(\ref{eq:Fx_solved}) and~(\ref{eq:Fy_solved})
assume constant values of frequency $\omega$ and decay rates
$\Gamma_{m+1,m}$. This is indeed valid if lasers are running cw and if
we scan the frequency $\omega$ slowly enough. But some experiments
must be carried out in a setting with pulsed lasers, e.g.~one might
wish to prepare the spin state in the maximally oriented state $F=4$,
$m_F = 4$ by illuminating atoms by a pulse of resonant, circularly
polarized laser light. Such state preparation has been used, e.g.~for
creation of entanglement between two samples of cesium
atoms~\cite{julsgaard:01}. For the magneto-optical resonance method to
be useful in such experiments it must be utilized in the correct
experimental conditions which now means time varying decay rates
$\Gamma_{m+1,m}$. In this section we outline the extensions into the
pulsed regimes and discuss the applicability of the magneto-optical
resonance method for characterization of spin states under these
circumstances.

We assume for simplicity that atoms are pumped to $m_F=4$ to an extent
that we only need to consider transitions between $m_F=3$ and $m_F=4$.
The extension to all levels should be straightforward (but
cumbersome). For these two levels we may write
Eq.~(\ref{eq:time_deriv_dens12}) as
\begin{equation}
\frac{\partial \slowdensop{34}}{\partial t}  
= (i\Delta - \Gamma/2)\slowdensop{34} 
+ i\chi[\densop{44}-\densop{33}] 
\end{equation}
where $\Delta = \omega - \omega_{43}$ and $\chi = g_F\mu_B
B_{RF}C(F,3)/4\hbar$. We assume as in
section~\ref{sec:solut-equat-moti} that the populations $\densop{44}$
and $\densop{33}$ can be treated as constants corresponding to small
angle deviations from the $z$-axis. Then the solution of the above
equation is straightforward
\begin{equation}
\begin{split}
\label{eq:solution_pulsed}
  \slowdensop{34}(t) &= \slowdensop{34}(0)e^{(i\Delta-\Gamma/2)t} \\
  &- \frac{i\chi}{i\Delta-\Gamma/2}[\densop{44}-\densop{33}]
  \left(1-e^{(i\Delta-\Gamma/2)t}\right)
\end{split}
\end{equation}
This solution starts out with $\slowdensop{34}(0)$ at $t=0$ and makes
a damped oscillation toward the steady state value
$-i\chi[\densop{44}-\densop{33}]/(i\Delta-\Gamma/2)$. Note, this
steady state value is exactly the result
in~(\ref{eq:solution_densop12}), and it is reached in a time
$\approx\Gamma^{-1}$. With the solution of $\slowdensop{34}$ we can
continue to find the actual spin, e.g.~$\F{x}$ given by~(\ref{Fx}) and
predict the results of a measurement.

Experimentally, we set up pumping lasers and a probe laser measuring
the transverse spin state as in Fig.~\ref{fig:exp_setup}b. The lasers
are turned on and off with acousto- and electro-optical modulators.
The decay rate in the absence of lasers is denoted
$\Gamma_{\mathrm{dark}}$ which is typically small. When the probe
laser is on an additional broadening of $\Gamma_{\mathrm{probe}}$ is
present leading to a total decay rate of
$\Gamma_{\mathrm{probe}}+\Gamma_{\mathrm{dark}}$. During the optical
pumping pulse the atoms are typically subject to a high decay rate
given in total by $\Gamma_{\mathrm{pump}}+\Gamma_{\mathrm{dark}}$.
The probe laser is typically turned on shortly after the optical
pumping has been turned off and is maintained for a time shorter than
the decay time $(\Gamma_{\mathrm{probe}}+\Gamma_{\mathrm{dark}})^{-1}$.
We are thus in the transient regime of Eq.~(\ref{eq:solution_pulsed})
and given the frequency $\omega$ of the driving RF-magnetic field we
cannot obtain a simple estimate of the amplitude of the response at
that frequency as in~(\ref{def:MORS}).  Instead we have time varying
quadrature components of the measured AC-signal and we simply
average these over time in the presence of the probe laser.  From the
perspective of modeling we need to evolve $\slowdensop{34}$ according
to~(\ref{eq:solution_pulsed}) with the relevant decay rates and
integrate the result over the time of the probe laser pulse.

An example of an experimental trace with corresponding numerical
modeling is shown in Fig.~\ref{fig:pulsed_graph}. The inset shows the
pulse sequence of lasers with a total period of 15ms. The repump and
optical pump lasers are turned on for 1.0ms with a high power making a
power broadening $\Gamma_{\mathrm{pump}} = 770\pm30$Hz. After a 0.3ms
period with lasers off the probe laser is fired for a 0.5ms period
giving rise to a small power broadening $\Gamma_{\mathrm{probe}}
\approx 2$Hz in addition to the background value around
$\Gamma_{\mathrm{dark}}\approx 18$Hz. These decay rates are estimated
from cw experiments carried out with the same laser powers. The
experimental trace matches quite well the calculated spectrum in the
sense that the general trend with a central peak and some broader
background structure is present. Also, the 15ms period together with
the (rotating frame) oscillation at frequency $\Delta$ introduces
ripples in both the experimental and calculated traces with spacing
$(15\mathrm{ms})^{-1} = 67\mathrm{Hz}$. However, there is some
asymmetry in the experimental trace which naturally is not present in
the modeling. This asymmetry could be an effect of non-perfect
orientation or maybe an effect of the frequency scan being a little
too fast. But this is a minor detail to the general impression that
the modeling is doing quite well.

%
% Data from 28-10-2002 and 07-01-2003 document the following.
%
We have tried to vary the duration of the pump pulse, the duration of
the probe pulse, and the value of $\Gamma_{\mathrm{pump}}$. In all
cases we maintain the agreement between measured and calculated
spectra as demonstrated for the single example in
Fig.~\ref{fig:pulsed_graph}. If $\Gamma_{\mathrm{pump}}$ is strong,
the magneto-optical resonance signal is essentially a broad feature
the structure of which in principle can be calculated. But varying the
total number of atoms we have seen experimentally that a simple
relation like $J_z \propto \sqrt {\mathrm{area}\cdot\mathrm{width}}$
practically holds very well where $J_z$ is measured by a DC-Faraday
rotation signal~(\ref{eq:DC-signal}) and the area and width refer to
the pulsed spectrum. This relation is exact in the cw case for $p=1$
but is apparently useful in some cases for the pulsed spectrum. This
is an advantage since we in these cases need not perform the more
cumbersome modeling suggested by~(\ref{eq:solution_pulsed}).

\section{Conclusion}
\label{sec:conclusion}
We have studied the magneto-optical resonance for cesium in a vapor
cell as a method for characterization of the quantum spin state of the
atomic vapor.  We have optically pumped the atoms to various
distributions among the Zeeman sublevels and exploited the
magneto-optical resonance signal, in particular the quadratic Zeeman
effect, to characterize these distributions and the decay rate of
coherences among the levels.

The theoretical description relies on well known techniques combined
with ad hoc tailored models for our particular needs. Specially useful
has been the relation $\densop{m,m} \propto \epsilon^m$ which just
happened to be a sufficient model in many of our cases. Our
experiments agree very well with the models in the cw case and
demonstrate that signals obtained in the pulsed regime are useful and
well understood. Our approach with the quadratic Zeeman effect is
particularly well suited for characterization of the ground spin state
within a hyperfine multiplet. Given the off-resonant laser probe with
low photon scattering rate we obtain the required high resolution. The
methods could be extended to the microwave region and cover
transitions between different hyperfine levels similar to the approach
in \cite{avila:87}. For excited spin state characterization the fast
spontaneous decay rules out high resolution methods. In this case
scattered photons from the pumping process may provide useful
information \cite{fischer:82}.

We have been able to create the coherent spin
state to a high degree of accuracy (better than 98\%) which is a very
good starting point for studies of quantum effects in our spin
ensemble. One limitation of our procedure is the fact that high
orientations and high pumping rates decrease the possible resolution
as discussed in section~\ref{sec:unresolved}. Thus it takes some
effort and experimental stability to see the difference between a 98\%
and a 100\% polarized sample. Also, a pulsed laser setup complicates
the conclusions but valuable information can still be extracted. The
use of 894nm pumping light on the $D1$-transition turns the $F=4$,
$m_F=4$ state into a dark state which has proved to be essential to
obtain high degrees of orientation.

The reliable method for characterization of a coherent spin state of
atoms in a paraffin coated vapor cell described above provides the
basis for implementation of various entanglement and quantum
communication protocols utilizing collective atomic spin states.
\begin{acknowledgments}
  This work was funded by the Danish National Research Foundation and
  the EU grants CAUAC and QUICOV.
\end{acknowledgments}
\appendix
\section{The Quadratic Zeeman Effect}
\label{sec:quadr-zeem-effect}
The quadratic Zeeman effect is well understood \cite{ramsey:56}, we
will just outline the important results for completeness of this
paper. An alkaline atom in an external magnetic field $\vec{B}$ is
described by the Hamiltonian
\begin{equation}
\label{eq:QZ_hamiltonian}
\H = h a \vec{I} \cdot \vec{J} - \frac{\mu_J}{J}\vec{J}\cdot\vec{B}
   - \frac{\mu_I}{I}\vec{I}\cdot\vec{B}
\end{equation}
where $\vec{J}$ describes the angular momentum of the outermost
electron, $\vec{I}$ is the nuclear spin, $a$ describes the strength of
the magnetic dipole interaction between the electronic and nuclear
spin, and $h$ is Planck's constant. The magnetic moment of the
electron (for an $s$-electron with $L=0$) is $\mu_J =
-1.0011596521869(41)\mu_B$. The value for the nuclear moment in cesium
is $\mu_I = 2.582025(4)\mu_N$. Thus, the last term
in~(\ref{eq:QZ_hamiltonian}) always gives a minor correction compared
to the second term, but the relative strength between the first and
second terms depends on the magnetic field $\vec{B}$.

The exact solution for the energy $E$ to the above Hamiltonian is
\begin{equation}
\label{eq:exact_energy}
E_{F,m} = -\frac{h\nu_{\mathrm{hfs}}}{2(2I+1)} - \frac{\mu_I}{I} B m
\pm \frac{h\nu_{\mathrm{hfs}}}{2}\sqrt{1+\frac{4m}{2I+1}x + x^2}
\end{equation}
where $\pm$ is used for $F=I\pm 1/2$, $m$ is the magnetic quantum
number (quantized along the direction of the magnetic field), $B =
|\vec{B}|$, and the hyperfine splitting $\nu_{\mathrm{hfs}}$ relates
to $a$ by $h\nu_{\mathrm{hfs}} = \frac{ha}{2}(2I+1)$.
The parameter $x$ describes the relative strength between the Zeeman
effect and the hyperfine splitting:
\begin{equation}
x = \frac{(-\mu_J/J + \mu_I/I)B}{h\nu_{\mathrm{hfs}}}
\end{equation}
For weak fields $m$ describes the projection of the total angular
momentum $\vec{F} = \vec{I}+\vec{J}$. The energy
levels~(\ref{eq:exact_energy}) can be seen in the inset of
Fig.~\ref{fig:intro_qz} as a function of the strength of the magnetic
field. We see that for small field strengths or very strong fields,
the energy depends linearly on $B$. In the intermediate region the
situation is quite non-linear. Our experiment is performed in the weak
field regime with $x \approx 3\cdot 10^{-4}$. Here a linear
approximation is very good, but it is important to calculate also the
second order contribution.

We study the magnetic sublevels by inducing magnetic transitions with
$\Delta m = \pm 1$. Thus, it will be interesting to calculate the
separation of adjacent sublevels.  We start out by
expanding~(\ref{eq:exact_energy}) to first order in the magnetic field
strength $B$ (leaving out the constant shift independent of $B$). With
the standard convention
\begin{equation}
\label{eq:E_zeeman_1st_order}
E_{F,m}^{(1)} = g_F \mu_B B m
\end{equation}
we get for cesium with nuclear spin $I = 7/2$ 
\begin{equation}
\label{eq:g_F}
\begin{split}
g_F &= \frac{1}{\mu_B}
\left(-\frac{\mu_I}{I} \pm \frac{-\mu_J/J+\mu_I/I}{2I+1}\right) \\
&= \left\{ 
\begin{split}
&0.250390 \quad\text{for } F=4 \\
-&0.251194 \quad\text{for } F=3
\end{split}
\right.
\end{split}
\end{equation}
These two numbers differ in magnitude by approximately 0.3\%. Thus, in
the lower linear regime of the inset of Fig.~\ref{fig:intro_qz} we
have a slightly higher separation between levels for the case of $F=3$
than $F=4$.

To calculate the quadratic Zeeman shift, it will suffice to do the
approximation $\mu_I = 0$. In this case we may write the first order
expansion of~(\ref{eq:exact_energy}) as $h\nu_{\mathrm{L}} \equiv
E_{m+1} - E_m = \frac{-\mu_J/J}{2I+1}\cdot B$, and we then easily
derive to second order
\begin{equation}
\frac{E_{m+1} - E_{m}}{h} = \nu_L\left( 1 -
\frac{\nu_\mathrm{L}}{\nu_{\mathrm{hfs}}}(2m+1)\right)
\end{equation}
This equation describes the transition frequency between the $m$'th
and the $(m+1)$'th level. The separation $\nu_{\mathrm{QZ}}$ caused by
the quadratic Zeeman effect between two \emph{lines} in
e.g.~Fig.~\ref{fig:intro_qz} will thus be
\begin{equation}
\label{def:QZ_splitting}
\nu_{\mathrm{QZ}} = \frac{2\nu_{\mathrm{L}}^2}{\nu_{\mathrm{hfs}}}
\end{equation}
All our experiments reported in this paper will have
$\nu_{\mathrm{L}}$ in the vicinity of 325kHz corresponding to a
magnetic field of a little less than 1 Gauss. With the cesium
hyperfine splitting being $\nu_{\mathrm{hfs}} = 9.1926$GHz we get a
quadratic Zeeman splitting of 23Hz.

\bibliographystyle{apsrev} \bibliography{bibfile}

\end{document}